\documentclass[twocolumn]{openjournal}

\usepackage{bm}
\usepackage{xspace}
\usepackage{xcolor}
\usepackage{amsmath}
\usepackage{graphicx}
\usepackage[hidelinks]{hyperref}
\hypersetup{breaklinks=true,colorlinks=true,linkcolor=blue,urlcolor=blue,citecolor=blue}
\usepackage[caption=false]{subfig}
\usepackage{booktabs}
\usepackage[capitalise]{cleveref}%must be added after amsmath
\usepackage{orcidlink}

\usepackage{savesym}
\savesymbol{tablenum}
\usepackage{siunitx}
\restoresymbol{SIX}{tablenum}

\usepackage{tabularx}

\usepackage{lineno}

\usepackage{ragged2e}

\newcommand{\figuredesc}[1]{% Figure description
  \begingroup
  \par
  \justifying\small
  \noindent #1
  \par
  \endgroup}

\creflabelformat{equation}{#2\textup{#1}#3}

\newcommand{\LCDM}{$\Lambda$CDM}

\newcommand*{\TT}{\ensuremath{T\!T}}
\newcommand*{\EE}{\ensuremath{E\!E}}
\newcommand*{\TE}{\ensuremath{T\!E}}

\newcommand*{\TTTEEE}{\ensuremath{T\!T/T\!E/E\!E}}

\newcommand*{\planck}{\textit{Planck}}

\newcommand*{\candl}{\texttt{candl}}

\newcommand{\RNum}[1]{\MakeUppercase{\romannumeral #1}}

\graphicspath{{./}{figures/}}

\defcitealias{dunkley13}{D13}

\shorttitle{Compressed 'CMB-lite' Likelihoods Using Automatic Differentiation}
\shortauthors{L. Balkenhol}

\begin{document}

\journalinfo{The Open Journal of Astrophysics}

\title{Compressed 'CMB-lite' Likelihoods Using Automatic Differentiation\vspace{-10ex}}

\author{L. Balkenhol\,\orcidlink{0000-0001-6899-1873}$^{1,\ast}$}
\email{$^\ast$lennart.balkenhol@iap.fr}

\affiliation{
$^1$Sorbonne Universit\'{e}, CNRS, UMR 7095, Institut d'Astrophysique de Paris, 98 bis bd Arago, 75014 Paris, France\\
}

\begin{abstract}
The compression of multi-frequency cosmic microwave background (CMB) power spectrum measurements into a series of foreground-marginalised CMB-only band powers allows for the construction of faster and more easily interpretable 'lite' likelihoods.
However, obtaining the compressed data vector is computationally expensive and yields a covariance matrix with sampling noise.
In this work, we present an implementation of the CMB-lite framework relying on automatic differentiation.
The technique presented reduces the computational cost of the lite likelihood construction to one minimisation and one Hessian evaluation, which run on a personal computer in about a minute.
We demonstrate the efficiency and accuracy of this procedure by applying it to the differentiable SPT-3G 2018 \TTTEEE{} likelihood from the \candl{} library.
We find good agreement between the marginalised posteriors of cosmological parameters yielded by the resulting lite likelihood and the reference multi-frequency version for all cosmological models tested;
the best-fit values shift by $<0.1\,\sigma$, where $\sigma$ is the width of the multi-frequency posterior, and the inferred parameter error bars match to within $<10\%$.
We publicly release the SPT-3G 2018 \TTTEEE{} lite likelihood and a python notebook showing its construction on the \href{https://github.com/Lbalkenhol/candl}{\candl{}} website.
\end{abstract}

\maketitle

%%%%%%%%%%%%%%%%%%%%%%%%%%%%%%%%%%%%%%%%%%%%%%%%%%

%%%%%%%%%%%%%%%%% BODY OF PAPER %%%%%%%%%%%%%%%%%%

\section{Introduction}\label{sec:intro}

Observations of the cosmic microwave background (CMB) anisotropies form a cornerstone of modern cosmology.
Data from recent and contemporary experiments, such as the \planck{} satellite \citep{planck18-1, planck18-6, planck18-5}, the South Pole Telescope (SPT) \citep{carlstrom11, dutcher21, balkenhol23, pan23, ge24}, the Atacama Cosmology Telescope (ACT) \citep{kosowsky03, aiola20, choi20, madhavacheril23, qu23}, and the BICEP/Keck experiments \citep{keating03b, staniszewski12, bicep2keck21} allow us to test the $\Lambda$ (cosmological constant) ($\mathrm{CDM}$) cold-dark-matter model across a wide range of angular scales in temperature and polarisation and search for new physics.

These experiments typically supply us with measurements of the CMB anisotropy power spectra at different observational frequencies binned into band powers;
from $N$ frequency channels $\geq N(N-1)/2$ multi-frequency spectra are constructed.
At the likelihood level, one then fits for the common CMB signal based on a cosmological model and the varying foreground contamination in the multi-frequency band powers along with any systematic effects.
The ensemble of foreground and systematic parameters is referred to as nuisance parameters.
While, as their name suggests, the nuisance parameters are not of primary interest, they must be included in a likelihood analysis to obtain the correct results on cosmological parameters.
However, nuisance parameters can be more numerous and hence slow down a Markov Chain Monte Carlo (MCMC) analysis by increasing the dimensionality of the parameter space to be explored and by slowing down the evaluation time of the likelihood due to the additional components in the data model.

While different methods have been proposed to compress the multi-frequency data into a shorter vector containing only the information of interest, the most common strategy for CMB power spectrum analyses is the construction of a so-called \emph{lite} likelihood, which contains a set of foreground-marginalised CMB-only band powers.
This allows for faster cosmological analyses primarily through the reduction of the number of nuisance parameters.
%At the same time, the covariance of the CMB-only band powers accounts for the combination of multi-frequency information as well as for the marginalisation over undesired signals.
%The lite likelihood hence has a shorter data vector and reduced number of nuisance parameters compared to the multi-frequency likelihood, greatly speeding up MCMC analyses.
The CMB-lite framework was first introduced by \citet{dunkley13} (hereafter \citetalias{dunkley13}) and has been applied to ACT (\citetalias{dunkley13}; \citet{choi20}), SPT \citep{calabrese13}, Planck \citep{planck15-11}, and most recently BICEP/Keck data \citep{prince24}.
Though lite likelihoods are advantageous once at hand, their construction is computationally expensive.
As set out by \citetalias{dunkley13}, it typically involves sampling a high-dimensional parameter space ($\mathcal{O}(10^2)$) with an alternating Gibbs-sampling technique.
This procedure naturally produces a covariance matrix with sampling noise, which must be reduced sufficiently or mitigated in another way to avoid adverse effects in subsequent cosmological analyses \citep{hartlap06, dodelson13, percival22, balkenhol22}.

Recently, the application of automatic differentiation in cosmology has seen a surge in popularity (see \citet{campagne23} for a series of example applications).
In this work, we show that by using the JAX-friendly \citep{jax18}, differentiable CMB likelihood library \candl{} \citep{balkenhol23} we can construct lite likelihoods in circa one minute on a personal computer and eliminate the sampling noise in the covariance matrix.
We apply this procedure to the SPT-3G 2018 \TTTEEE{} data set \citep{balkenhol23} and benchmark the performance of the lite likelihood against its multi-frequency counterpart.
The lite likelihood and code to construct CMB-only band powers is made publicly available on the \candl{} website.\footnote{\url{https://github.com/Lbalkenhol/candl}}

We begin with background on the CMB-lite framework and the \candl{} library in Section \ref{sec:background}.
In Section \ref{sec:technique} we outline our technique for the construction of CMB-lite likelihoods and subsequently apply it to the SPT-3G 2018 \TTTEEE{} data set in Section \ref{sec:app}.
We share our conclusions in Section \ref{sec:conclusions}.

\section{Background}\label{sec:background}

We briefly review the basics of CMB likelihood inference and outline the CMB-lite framework as introduced by \citetalias{dunkley13}.
A CMB power spectrum analysis typically yields multi-frequency band powers $C^{\mu\nu,\mathrm{data}}_b$, i.e. the binned cross power spectrum between observations in frequency bands $\mu$ and $\nu$.
The weights of individual multipole moments of the power spectrum $\ell$ in a band power bin $b$ are given by the window function $W_{b\ell}$.
We usually also have access to the covariance of the multi-frequency band powers, $\Sigma_{bb'}$.
We index binned band powers with subscripts $b$ and $b'$, the unbinned power spectrum with $\ell$, and use repeated indices to imply summation.

We formulate model band powers $C_b^{\mu\nu,\mathrm{model}}(\phi)$ to describe the data, where the $\phi$ parameters contain CMB band powers $C_b^{\mathrm{CMB}}$, as well as nuisance parameters $\theta$, $\phi=(C_b^{\mathrm{CMB}}, \theta)$.
The nuisance parameters enter a frequency-dependent foreground model $C_b^{\mu\nu,\mathrm{FG}}(\theta)$ and calibration factors $y^{\mu\nu}(\theta)$.
Explicitly, our \emph{data model} is:
\begin{align}
\label{eq:lite_model}
    C_b^{\mu\nu,\mathrm{model}}(\phi) = y^{\mu\nu}(\theta) \bigg( &A^{\mu\nu} C_b^{\mathrm{CMB}} \nonumber \\
    &+ C_b^{\mu\nu,\mathrm{FG}}(\theta) \bigg).
\end{align}
Here, $A^{\mu\nu}$ is sometimes referred to as the mapping-matrix \citepalias{dunkley13} or the design-matrix \citep{mocanu19, planck18-5, dutcher21, balkenhol23}; this matrix of zeros and ones connects the multi-frequency estimates of the same CMB band power in the data vector to the corresponding single element in $C_b^{\mathrm{CMB}}$.
Finally, to perform a cosmological analysis we must assume a cosmological model, i.e. a prescription that produces predictions for the CMB power spectrum $C^{\mathrm{CMB}}_{\ell}(\psi)$ for a set of cosmological parameter values $\psi$, which we can then bin to obtain $C_b^{\mathrm{CMB}}(\psi) = W_{b\ell}C^{\mathrm{CMB}}_{\ell}(\psi)$.

The comparison of data and model is typically done via a Gaussian likelihood of the form
\begin{equation}
\label{eq:like_gauss}
    2 \ln{\mathcal{L}} \propto \Delta_{b} \Xi_{bb'}^{-1} \Delta_{b'},
\end{equation}
where $\Delta$ represents the difference between data and model and $\Xi$ an appropriate covariance matrix.
This general form finds different applications in CMB analyses.
The most straightforward one is to analyse our multi-frequency band powers with our data model assuming a cosmological model.
For this, we set $\Delta_{b}(\psi,\theta) = C^{\mu\nu,\mathrm{data}}_b - C_b^{\mu\nu,\mathrm{model}}(C_b^{\mathrm{CMB}}(\psi),\theta)$ along with $\Xi_{bb'} = \Sigma_{bb'}$.
We refer to this case as the \emph{multi-frequency likelihood}, $\mathcal{L}^{\mu\nu}$.
Typically, one adds prior knowledge on $\psi$ and $\phi$ to the multi-frequency likelihood and explores the resulting posterior with an MCMC approach.
This is a Bayesian analysis and yields marginalised constraints on cosmological and nuisance parameters alike.

In contrast, the CMB-lite framework first seeks to determine the common CMB signal in the multi-frequency band powers, i.e. $C_b^{\mathrm{CMB}}$, without assuming a cosmological model, marginalising over nuisance parameters in the process.\footnote{Note that in the CMB-lite framework the term \emph{secondary parameters} is occasionally used to describe parameters that are marginalised over during the construction of the CMB-only band powers. So while the label nuisance parameters refers to everything non-cosmological, secondary parameters are only the subset of nuisance parameters not wanted in the final lite likelihood.
We assume these are equivalent for the sake of simplicity here, though as we will see in Section \ref{sec:app} it can be beneficial to retain some select nuisance parameters in the lite likelihood.}
In order to do so one samples over $\phi$, treating the CMB band powers as parameters to be explored, and compares the model prediction to the measured data by setting $\Delta_{b}(\phi) = C^{\mu\nu,\mathrm{data}}_b - C_b^{\mu\nu,\mathrm{model}}(\phi)$ and $\Xi_{bb'} = \Sigma_{bb'}$ in Equation \ref{eq:like_gauss}.
We refer to this likelihood as the \emph{reconstruction likelihood}, $\mathcal{L}^{\mathrm{recon}}$ and its exploration, where we may also consider any priors on $\phi$, as the \emph{reconstruction process}.
The reconstruction process thus yields estimates of the CMB power in the band power bins, $C_b^{\mathrm{lite}}$, as well as their covariance $\Sigma_{bb'}^{\mathrm{lite}}$, which contains a suitable contribution for the nuisance parameters.
In \citetalias{dunkley13}, the reconstruction likelihood is analysed with an MCMC approach by alternating Gibbs sampling steps for the CMB band powers and the nuisance parameters.

Once the reconstruction process is complete, a cosmological analysis is carried out in a second step.
Assuming a cosmological model, we set $\Delta_{b}(\psi) = C_b^{\mathrm{lite}} - C_b^{\mathrm{CMB}}(\psi)$ and $\Xi_{bb'} = \Sigma_{bb'}^{\mathrm{lite}}$ in Equation \ref{eq:like_gauss}, which yields the \emph{lite likelihood}, $\mathcal{L}^{\mathrm{lite}}$.
We can now once again carry out a Bayesian analysis to obtain marginalised posterior constraints on cosmological parameters $\psi$.
For the same model, the resulting distributions generally agree well with the ones obtained from the multi-frequency likelihood.
Still, small differences do arise, as during the reconstruction process one picks up fluctuations from noise and foreground contamination and assigns these to the CMB-only band powers.
In the lite likelihood these may project differently onto cosmological parameters compared to the multi-frequency likelihood, where cosmological and nuisance parameters are explored simultaneously.
Typically though, this realisation-dependent bias is below or comparable to the level of numerical noise expected in parameter estimation, which we take to be shifts in central values of marginalised posteriors by $10\%$ of their width.

There are numerous advantages to the CMB-lite framework.
First, it reduces the size of the data vector, speeding up likelihood evaluation.
It also eases the exploration of the parameter space by reducing its dimensionality.
Moreover, it combines the common information from the multi-frequency band powers in an optimal way, forming a minimum-variance combination of the data when the CMB signal is dominant.
At the same time, uninteresting information is marginalised over and the resulting uncertainty is neatly captured in the covariance of the CMB-only band powers.
This makes lite likelihoods also more intuitively interpretable.
Crucially, no cosmological model enters the reconstruction likelihood.
This means the reconstruction procedure does not introduce additional model-dependence beyond what already exists in the reference multi-frequency likelihood.

However, the framework is not without downsides.
If cosmological parameters are strongly correlated with nuisance parameters, constraints from the multi-frequency and lite likelihoods may differ non-negligibly.
On a technical level, the MCMC analysis of the reconstruction likelihood is computationally expensive and leads to sampling noise in the covariance matrix estimate, which can cause adverse effects in the parameter constraints of the lite likelihood if not reduced sufficiently or otherwise mitigated \citep{hartlap06, dodelson13, percival22, balkenhol22}.
Of course, one is always free to run the reference multi-frequency likelihood when in doubt, though having to do so continuously diminishes the value of the CMB-lite framework.

In this work, we ameliorate the numerical challenges associated with the reconstruction procedure.
To do so, we use the \candl{} library of CMB likelihoods, which allows for easy, intuitive access to data from the SPT and ACT collaborations and, crucially, is written in a JAX-friendly way \citep{jax18}.
JAX is a google-developed python package which exposes the code to an automatic differentiation algorithm that allows for the accurate and fast calculation of derivatives of \candl{} likelihoods without relying on finite difference methods.\footnote{\url{https://github.com/google/jax}}

\begin{figure*}[ht!]
    \includegraphics[width=2.0\columnwidth]{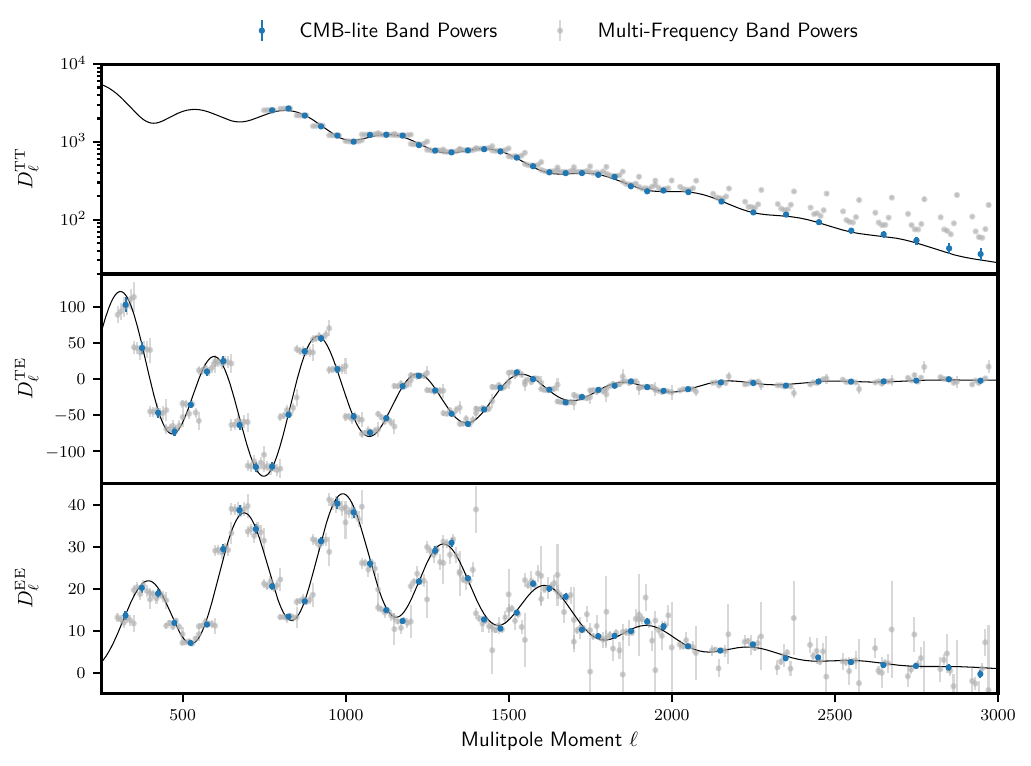}
    \caption{Foreground-marginalised CMB-only band powers (blue) based on the SPT-3G 2018 \TTTEEE{} multi-frequency data (grey).
    Though the construction of the CMB-only data vector does not require a cosmological model, the data visually follow the best-fit \LCDM{} model of the multi-frequency likelihood shown in black.
    The CMB-lite band powers visibly remove the foreground contamination in the temperature power spectrum.
    }
    \label{fig:bdp}
\end{figure*}

\section{Technique}\label{sec:technique}

Thanks to \candl{} and JAX the reconstruction likelihood is differentiable, meaning we can evaluate its first- and second-order derivates
\begin{equation}
    \left. \frac{\partial\mathcal{L}^{\mathrm{recon}}}{\partial\phi_i}\right|_{\phi}, \hspace{0.4cm} \left( \frac{\partial^2\mathcal{L}^{\mathrm{recon}}}{\partial\phi_i\partial\phi_j}\right)_{\phi}
\end{equation}
at any desired point $\phi$ quickly and accurately.
Access to the gradient allows us to find the best-fit parameters of the reconstruction likelihood extremely fast using, e.g. the truncated Newton or Newton-Raphson methods \citep{nash84, nocedal06}.
The Hessian is useful because evaluated at the best-fit point it is equal to the negative Fisher matrix and hence connects to the covariance of the CMB-only band powers \citep{heavens14}.

With access to these functions, we perform the reconstruction as follows:
\begin{enumerate}
    \item Minimise the reconstruction likelihood, finding its best-fit point. This yields the CMB-only band powers.
    \item Evaluate the Hessian of the reconstruction likelihood at the best-fit point. The negative inverse of this matrix serves as the covariance of the CMB-only band powers.
\end{enumerate}
Hence, we obtain all necessary products for the lite likelihood from minimising the reconstruction likelihood once and one evaluation of its Hessian.
The above prescription greatly reduces the computational cost of the reconstruction procedure such that it can be performed on a personal laptop in about a minute.
This means one can easily perform multiple reconstructions to assess robustness, using for example different foreground models or angular scale cuts.
Moreover, this procedure eliminates numerical noise from the covariance matrix of CMB-only band powers.
Though automatic-differentiation also has its limits, second-order derivatives are usually well behaved and we have not seen any suggestive artefacts in the application to follow in Section \ref{sec:app} \citep{campagne23}.

We note that while in principle the use of the Hessian of the reconstruction likelihood for the covariance of the CMB-only band powers does not capture non-Gaussian contributions, we did not detect significant biases due to this during the application to SPT data in Section \ref{sec:app}.
One can test for this effect by reconstructing a known scatter-free model spectrum and comparing parameter constraints for models of interest between the resulting lite and multi-frequency likelihoods.
Should differences turn out to be sizeable, one may consider the methods put forward by \citet{sellentin14} or \citet{schaefer22} in order to improve the covariance estimate or resort to MCMC sampling the reconstruction likelihood.
For this the use of gradient-based sampling methods (e.g. Hamiltonian Monte Carlo, No U-Turns) is also promising as these tend to perform well in high-dimensionality scenarios \citep{hoffman11}.

\section{Application to Data}\label{sec:app}

In this section we apply the above procedure to the SPT-3G 2018 \TTTEEE{} likelihood implemented in \candl{}.
The minimisation of the reconstruction likelihood is carried out using the truncated Newton algorithm implemented in scipy \citep{nash84, nocedal06, scipy}.
To benchmark the performance of the resulting lite likelihood and judge the success of the reconstruction procedure more generally we will consider three cosmological models: \LCDM{}, \LCDM{}+$N_{\mathrm{eff}}$, and \LCDM{}+$A_{L}$.
We parametrise these models using the Hubble constant $H_0$, the baryon and cold-dark-matter densities $\Omega_{\mathrm{b}} h^2$ and $\Omega_{\mathrm{c}} h^2$, respectively, the amplitude $\ln(10^{10} A_{\mathrm{s}})$ and spectral tilt $n_{\mathrm{s}}$ of the power spectrum of initial scalar fluctuations, the optical depth to reionisation $\tau$, plus optionally either the effective number of relativistic degrees of freedom $N_{\mathrm{eff}}$ or the amplitude of the gravitational lensing effect on the power spectrum $A_L$.
To calculate CMB spectra, we utilise the CosmoPower models trained for the analysis of the SPT-3G 2018 \TTTEEE{} data set on high-accuracy CAMB spectra \citep{lewis00, spuriomancini22, balkenhol23, piras23}.
To perform MCMC analyses of the multi-frequency and lite likelihoods in the aforementioned cosmological models we use Cobaya \citep{torrado21} with a Metropolis-Hastings sampler.
We use dedicated minimisers to obtain best-fit points, again using scipy's truncated Newton algorithm.
For cosmological analyses, we apply the same \planck{}-based Gaussian prior on $\tau$ centred on $0.054$ with width $0.0074$ as \citet{balkenhol23} in the lite likelihood \citep{planck18-6}.

The SPT-3G 2018 \TTTEEE{} multi-frequency likelihood has various noteworthy aspects for the reconstruction process.
First, the window functions differ across frequencies owing to the instrumental beam and the filter strategy of the analysis, which as \citet{prince24} points out means that there exists no single set of uniquely defined CMB-only band powers to reconstruct; in principle, one would have to assign a CMB-only band power bin to each window function shape.
%in the same spirit as \citetalias{dunkley13}; \citet{choi20, prince24}
However, since differences are small, we ignore them in the reconstruction procedure and use a combination of the multi-frequency window functions weighted by the band power covariance matrix in the lite likelihood.

Second, the multi-frequency likelihood involves the addition of a model- and frequency-dependent beam contribution to the band power covariance matrix.
To avoid having to represent this operation at the level of the lite likelihood, one could reformulate the multi-frequency likelihood to account for the uncertainty of the beam measurement through modifications of the model spectra controlled by additional nuisance parameters to be marginalised over during the reconstruction process \citep[see e.g.][]{henning18}.
However, we find that accounting for this term in the reconstruction likelihood and simply ignoring it in the lite likelihood is sufficiently accurate for the data set at hand.

\begin{figure}[hb!]
    \centering
    \includegraphics[width=\columnwidth]{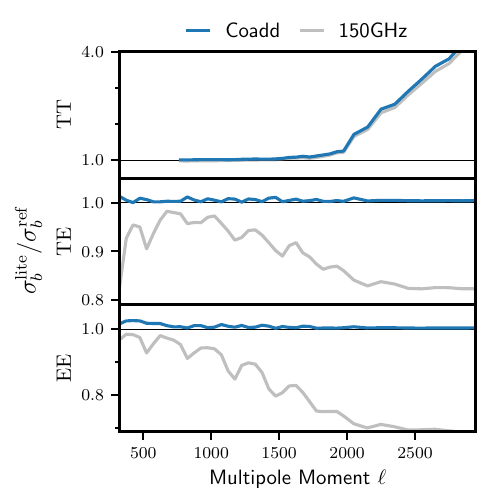}
    \caption{Ratio of the error bars of the lite likelihood to two reference cases ($\sigma^{\mathrm{ref}}_b$): the coadd (blue) and the $150\,\mathrm{GHz}$ auto-spectrum (grey).
    As foregrounds are already marginalised over in the CMB-lite band powers, error bars are larger than the two reference cases at high $\ell$ for \TT{}.
    For \TE{} and \EE{} we see that the lite likelihood improves on the $150\,\mathrm{GHz}$ auto-spectrum on all scales due to the combination of multi-frequency information.
    The beam and calibration uncertainties as well as the super-sample lensing effect cause an increase of $<3\%$ of the lite error bars with respect to the coadd for \TE{} and \EE{}.
    }
    \label{fig:err}
\end{figure}

\begin{figure*}[ht!]
    \centering
    \includegraphics[width=2.0\columnwidth]{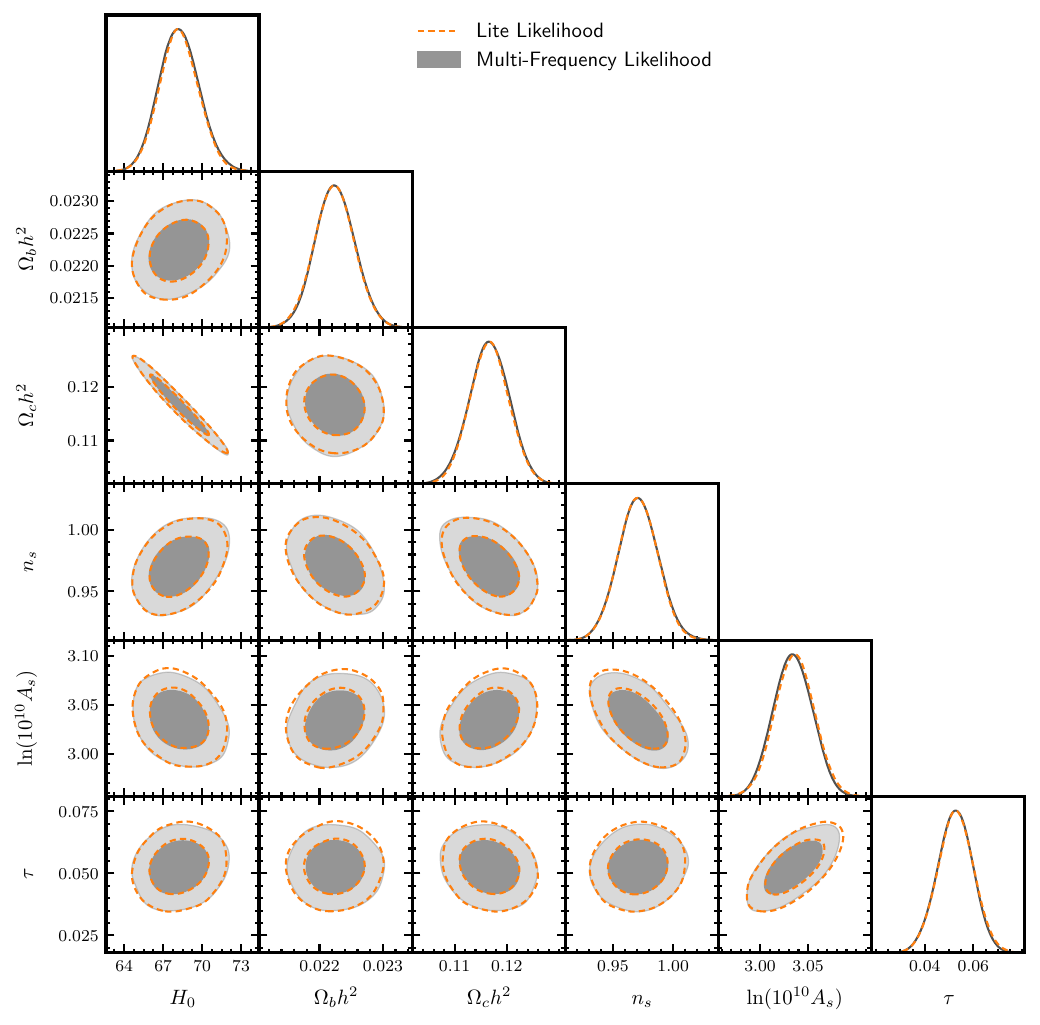}
    \caption{Marginalised posteriors for \LCDM{} parameters obtained from the full multi-frequency (grey filled contours) and lite (orange dashed line contours) SPT-3G 2018 \TTTEEE{} likelihoods ($68\%$ and $95\%$ confidence levels).
    The constraints match well, with only a small offset in $\ln{(10^{10}A_s)}$ of $<10\%$ of the width of the multi-frequency posterior visually discernable.}
    \label{fig:lcdm_triangle}
\end{figure*}

\begin{figure*}[ht!]
    \centering
    \includegraphics[width=2.0\columnwidth]{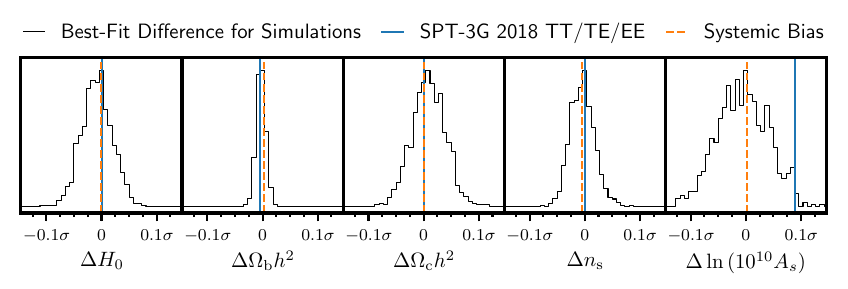}
    \caption{The difference in best-fit \LCDM{} parameters of the lite likelihood compared to the multi-frequency likelihood for 1000 mock band power realisations (black histogram) normalised by the width of the constraints of the multi-frequency likelihood ($1\sigma$).
    We indicate the bias we observe for the SPT-3G 2018 \TTTEEE{} data as blue vertical lines.
    Dashed orange vertical lines indicate the systemic bias described in Section \ref{sec:app}, which is subdominant.
    In general, with $\lesssim 0.1\,\sigma$, the size of the realisation-dependent scatter is at an acceptable level and even in bad cases comparable to the uncertainty inherent in an MCMC analysis.
    For the SPT-3G 2018 \TTTEEE{} data set, the only parameter with a noticeable shift is $\ln{(10^{10} A_s)}$, though the observed value is within $2.1$ standard deviations of the mean and hence statistically normal.}
    \label{fig:sims}
\end{figure*}

Third, we follow the suggestion of \citetalias{dunkley13} and split the calibration module of the multi-frequency likelihood into two: an internal calibration, relative to the $150\,\mathrm{GHz}$ \TT{} and \EE{} auto-spectra, and an external, absolute calibration of all spectra controlled by new parameters $T_{\mathrm{cal}}$ and $E_{\mathrm{cal}}$, which modify spectra like their frequency-dependent counterparts in the original likelihood.
By keeping $T_{\mathrm{cal}}$ and $E_{\mathrm{cal}}$ in the lite likelihood we avoid introducing significant long-range correlations between the estimated CMB-only band powers, which ultimately makes it easier to interpret residuals of the data vector.

Fourth, the transformation of model spectra by the aberration effect due to the Earth's relative motion to the CMB depends on the derivative of the unbinned CMB power spectrum with respect to the multipole moment $\ell$ \citep{jeong14}.
However, we are only interested in reconstructing CMB band powers, i.e. bins in multipole moments, and there exists no unique correspondence between the aberration effect at the band power and the unbinned CMB power spectrum level.
Therefore, we consider aberration as part of the signal and aim to encapsulate it in our reconstruction, also accounting for this effect in the lite likelihood.

Lastly, we face a similar issue for the super-sample lensing effect \citep{manzotti14}.
Though we could treat this phenomenon similarly to aberration, we would have to carry the parameter controlling the strength of the effect, the mean lensing convergence across the survey field, into the lite likelihood.
Instead, we add a suitable contribution to the band power covariance matrix and therefore keep the number of nuisance parameters and transformations in the lite likelihood low.

Together, these points lead to a small systemic bias, which we quantify by generating scatter-free mock band powers for a set of fiducial parameters in \LCDM{}.
We perform the reconstruction and compare constraints on \LCDM{} parameters between the multi-frequency and the lite likelihood.
The best-fit values of the lite likelihood are within $<0.01\,\sigma$ of the multi-frequency likelihood for all parameters, where $\sigma$ refers to the width of the marginalised posteriors yielded by the multi-frequency likelihood.
The error bars on cosmological parameters match to better than $1\%$.
Hence any inaccuracies or biases in the reconstruction procedure are negligible compared to the level of uncertainty ascribed to MCMC analyses and, as we demonstrate later on, to the realisation-dependent bias picked up by the reconstruction from noise and sample variance fluctuations when performed on real data.

We now perform the reconstruction procedure on the SPT-3G 2018 \TTTEEE{} data set.
The resulting CMB-only band powers are shown in Figure \ref{fig:bdp}.
Like their multi-frequency counterpart, the band powers cover the multipole ranges $750 \leq \ell < 3000$ in \TT{} and $300 \leq \ell < 3000$ in \TE{} and \EE{} with bin widths $\Delta\ell=50$ for $\ell < 2000$ and $\Delta\ell=100$ for $\ell>2000$.
Visually, we see that the power on small angular scales in temperature is reduced compared to the multi-frequency band powers as the foreground contamination is removed.

In Figure \ref{fig:err} we compare the error bars of the CMB-only band powers to the ones of the \emph{coadd} of multi-frequency spectra, i.e. the minimum-variance combination of the full data vector after foreground-subtraction \citep{planck15-11}.
Since the covariance of the coadd only reflects the combination of multi-frequency information to reduce noise, this comparison allows us to quantify the information loss due to foregrounds and systematic effects at the band power level.
Note that while the CMB-lite reconstruction does not require a cosmological model, this is not true for the coadd and we construct it using the best-fit point of the multi-frequency likelihood in \LCDM{}.
As expected, we see that the foreground contamination in temperature data leads to a gradual increase of the error bars in the lite likelihood with respect to the coadd towards small angular scales with the uncertainty on the final bin of the \TT{} spectrum being a factor of $4.4$ larger.
For the polarisation spectra on the other hand, the lite likelihood's error bars are inflated by $<3\%$ more evenly across angular scales;
this small increase reflects the uncertainties of the beam measurement and the calibration, as well as the expected size of super-sample lensing fluctuations for the survey field.
Though neighbouring band powers in the lite likelihood remain mildly correlated due to flat-sky projection effects \citep{balkenhol23}, we do not introduce any significant new long-range correlations owing to our treatment of calibration.

%\begin{figure*}
%    \centering
%    \includegraphics[width=2.0\columnwidth]{figures/error_ratio.pdf}
%    \caption{Error ratio.}
%    \label{fig:err_ratio}
%\end{figure*}

%We show the covariance matrix of the reconstructed CMB-only band powers and secondary parameters in Figure \ref{fig:cov}.
%The matrix appears as expected: showing a correlation between temperature foreground parameters and band powers already pointed out by \citetalias{dunkley13}.
%Close inspection also reveals broad, though relatively low correlations with the internal calibration parameters.

%\begin{figure}
%    \centering
%    \includegraphics[width=\columnwidth]{figures/full_cov.pdf}
%    \caption{full covmat. Hiding 150 relative calibration parameters that arent really real.}
%    \label{fig:cov}
%\end{figure}

We construct the lite likelihood based on the shown foreground-marginalised CMB-only band powers and their covariance.
While we still need to account for the effects of aberration and absolute calibration at the likelihood level as mentioned above, we have reduced the number of transformations in the data model from $11$ to two and the number of nuisance parameters from $33$ to two ($T_{\mathrm{cal}}$ and $E_{\mathrm{cal}}$) \citep[for details on the multi-frequency likelihood see \S \RNum{3}B and Table \RNum{8} of][]{balkenhol23}.
The length of the data vector has been reduced from $728$ to $123$ bins.
We find that the evaluation time of the lite likelihood (starting from pre-calculated CMB-spectra and a set of nuisance parameter values) is about a factor of four smaller than its multi-frequency counterpart and about half the evaluation time of the CosmoPower emulators, as measured by Cobaya.

The marginalised posteriors obtained from MCMC analyses of the two likelihoods in \LCDM{} are shown in Figure \ref{fig:lcdm_triangle}.
The constraints match very well visually.
This is also quantitatively true;
we tabulate the difference in best-fit points and the ratio of parameter errors for all models considered in Table \ref{tab:par_table}.
All parameter differences are below $0.1\,\sigma$, where $\sigma$ is the width of the marginalised parameter posteriors of the multi-frequency likelihood.
Furthermore, the inferred parameter error bars match to within $<10\%$ in all cases.
As \citet{planck18-5} found, we also see a slight degradation in the performance for the two standard model extensions considered ($N_{\mathrm{eff}}$ and $A_L$), as these models allow for changes to the damping tail of the power spectra, which leads to correlations with foreground parameters.
Still, the performance remains acceptable and matches what was achieved by \citetalias{dunkley13} and \citet{planck18-5}.
We conclude that given the numerical noise present in MCMC analyses the lite likelihood performs well.

\begin{table*}
    \caption{}
    \figuredesc{Comparison of best-fit points and the width of marginalised posteriors between the SPT-3G 2018 \TTTEEE{} multi-frequency and lite likelihoods for \LCDM{}, \LCDM{}+$N_{\mathrm{eff}}$, and \LCDM{}+$A_{L}$ models.
    The first column for each model lists the shift of best fit points obtained from dedicated minimiser runs normalised by the parameter error bars obtained from the MCMC analysis of the multi-frequency likelihood.
    The second column indicates the ratio of posterior widths.
    Still, all observed shifts are $<10\%$ the width of the posteriors obtained from the multi-frequency likelihood and the parameter errors match to within $10\%$, indicating good agreement between the two likelihoods.}
    \vspace{2ex}
    \label{tab:par_table}
\def\arraystretch{1.5}
    \centering
    \begin{tabular}{c S[table-format=2.3] S[table-format=2.3] c S[table-format=2.3] S[table-format=2.3] c S[table-format=2.3] S[table-format=2.3]}
    & \multicolumn{2}{c}{\LCDM{}} & & \multicolumn{2}{c}{\LCDM{}$+N_{\mathrm{eff}}$} & & \multicolumn{2}{c}{\LCDM{}$+A_L$} \\
    \cline{2-3}\cline{5-6}\cline{8-9}
    Parameter & {$\Delta\mu/\sigma^{\mu\nu}$} & {$\sigma^{\mathrm{lite}}/\sigma^{\mu\nu}-1$} & & {$\Delta\mu/\sigma^{\mu\nu}$} & {$\sigma^{\mathrm{lite}}/\sigma^{\mu\nu}-1$} & & {$\Delta\mu/\sigma^{\mu\nu}$} & {$\sigma^{\mathrm{lite}}/\sigma^{\mu\nu}-1$} \\
    \hline \hline
$H_0$ & 0.000 & -0.023 & & 0.022 & 0.016 & & -0.016 & 0.014 \\
$\Omega_{\mathrm{b}} h^2$ & -0.004 & 0.006 & & 0.015 & 0.021 & & -0.008 & -0.002 \\
$\Omega_{\mathrm{c}} h^2$ & 0.001 & -0.021 & & 0.020 & 0.059 & & 0.016 & 0.014 \\
$n_{\mathrm{s}}$ & 0.001 & -0.018 & & 0.018 & 0.017 & & -0.009 & 0.026 \\
$\ln(10^{10} A)$ & 0.089 & 0.020 & & 0.090 & 0.002 & & 0.082 & 0.017 \\
$\tau$ & -0.003 & 0.026 & & -0.001 & -0.008 & & 0.000 & 0.003 \\
$N_{\rm eff}$ & {--} & {--} & & 0.023 & 0.040 & & {--} & {--} \\
$A_{\rm L}$ & {--} & {--} & & {--} & {--} & & -0.019 & -0.011 \\
\hline
\hspace{0.01cm}
    \end{tabular}
\end{table*}

Still, we want to reassure ourselves that the shifts observed are compatible with the expectation from the realisation-dependent bias.
Hence, we run 1000 simulations in which we draw Gaussian band power realisations from the multi-frequency band power covariance matrix based on a fiducial \LCDM{} cosmology (with parameter values fixed to the results of \citet{planck18-6}).
We assign these band powers to the reference multi-frequency likelihood, minimise it, run the reconstruction procedure on them, and minimise the resulting lite likelihood, assuming the \LCDM{} model where appropriate.
We then compare the difference in the best-fit parameters obtained from the multi-frequency and lite likelihoods for each realisation; the corresponding distribution is shown in Figure \ref{fig:sims}.
The realisation-dependent bias is typically confined to within $\pm 0.1\,\sigma$, where $\sigma$ is the width of the parameter posterior obtained from the multi-frequency likelihood, signalling the good robustness of the procedure.
The specific reconstruction for the SPT-3G 2018 \TTTEEE{} data is remarkably bias free for $H_0$, $\Omega_{\mathrm{b}} h^2$, $\Omega_{\mathrm{c}} h^2$ and $n_{\mathrm{s}}$.
The bias seen for $\ln{(10^{10}) A_s}$ is consistent with zero within two standard deviations of the distribution obtained from simulations and compatible with a statistical fluctuation.

Lastly, we check that the value of the reconstruction likelihood at its best-fit point, i.e. its minimum, is consistent with the corresponding distribution obtained from simulations.
This is the case at $1.2$ standard deviations from the mean.
This verifies that the data model chosen describes the measured band powers well.
We therefore conclude that the reconstruction procedure has been successful and any differences in parameter constraints between the SPT-3G 2018 \TTTEEE{} multi-frequency and lite likelihoods are consistent with the expected size of statistical fluctuations.
We make the lite version of the likelihood publicly available on the \candl{} website.

\section{Conclusions}\label{sec:conclusions}

In this work we have presented an implementation of the CMB-lite framework using automatic differentiation and have applied it to the differentiable SPT-3G 2018 \TTTEEE{} likelihood available in the \candl{} library.
The reconstruction runs in about a minute on a personal computer and yields a sampling noise-free covariance of the foreground-marginalised CMB-only band powers.
The resulting lite likelihood performs well and recovers the parameter constraints of the reference multi-frequency likelihood for \LCDM{}, \LCDM{}+$N_{\mathrm{eff}}$, and \LCDM{}+$A_{L}$ models to within good accuracy;
best-fit points are offset by $<0.1\,\sigma$, where $\sigma$ is the width of the marginalised posterior of the multi-frequency likelihood, and the inferred parameter error bars match to within $<10\%$ in all cases.
The lite likelihood is publicly available on the \candl{} website along with a python notebook demonstrating its construction.

More broadly, this work represents another successful application of automatic differentiation in cosmology.
This technique continues to increase in popularity in the community and not only facilitates improvements of existing methodologies as is the case here, but also enables entirely novel analyses.
This is a trend we expect to continue; as cosmological data continue to improve, so must our tools.
Looking ahead, Simons Observatory \citep{ade19} and CMB-S4 data \citep{abazajian19} will have broad frequency coverage to confidently separate cosmological and foreground signals.
This increases the number of multi-frequency spectra and hence also the advantages the CMB-lite framework can bring.
At the same time, the low noise levels of these experiments will compel many robustness tests, which are efficiently carried out using the numerical techniques presented in this work.

\begin{acknowledgments}
L. B. is deeply grateful to Silvia Galli for her encouragement to pursue the work, helpful conversations, and useful suggestions.
L. B. is also grateful to Karim Benabed for enlightening discussions and constructive suggestions.
L. B. would like to thank the following people for providing comments on the manuscript: Silvia Galli, Karim Benabed, and \'Etienne Camphuis.
This work uses JAX \citep{jax18} and the scientific python stack \citep{jones01, hunter07, vanDerWalt11}.
This project has received funding from the European Research Council (ERC) under the European Union’s Horizon 2020 research and innovation programme (grant agreement No 101001897).
This work has received funding from the Centre National d’Etudes Spatiales and has made use of the Infinity Cluster hosted by the Institut d’Astrophysique de Paris.

\end{acknowledgments}

%%%%%%%%%%%%%%%%%%%% REFERENCES %%%%%%%%%%%%%%%%%%

\bibliographystyle{aa}
\typeout{}
\bibliography{CMBlite}

\begin{thebibliography}{46}
\expandafter\ifx\csname natexlab\endcsname\relax\def\natexlab#1{#1}\fi

\bibitem[{{Ade} {et~al.}(2021){Ade}, {Ahmed}, {Amiri}, {Barkats}, {Thakur}, {Bischoff}, {Beck}, {Bock}, {Boenish}, {Bullock}, {Buza}, {Cheshire}, {Connors}, {Cornelison}, {Crumrine}, {Cukierman}, {Denison}, {Dierickx}, {Duband}, {Eiben}, {Fatigoni}, {Filippini}, {Fliescher}, {Goeckner-Wald}, {Goldfinger}, {Grayson}, {Grimes}, {Hall}, {Halal}, {Halpern}, {Hand}, {Harrison}, {Henderson}, {Hildebrandt}, {Hilton}, {Hubmayr}, {Hui}, {Irwin}, {Kang}, {Karkare}, {Karpel}, {Kefeli}, {Kernasovskiy}, {Kovac}, {Kuo}, {Lau}, {Leitch}, {Lennox}, {Megerian}, {Minutolo}, {Moncelsi}, {Nakato}, {Namikawa}, {Nguyen}, {O'Brient}, {Ogburn}, {Palladino}, {Prouve}, {Pryke}, {Racine}, {Reintsema}, {Richter}, {Schillaci}, {Schwarz}, {Schmitt}, {Sheehy}, {Soliman}, {Germaine}, {Steinbach}, {Sudiwala}, {Teply}, {Thompson}, {Tolan}, {Tucker}, {Turner}, {Umilt{\`a}}, {Verg{\`e}s}, {Vieregg}, {Wandui}, {Weber}, {Wiebe}, {Willmert}, {Wong}, {Wu}, {Yang}, {Yoon}, {Young}, {Yu}, {Zeng}, {Zhang}, {Zhang}, \& {Bicep/Keck
  Collaboration}}]{bicep2keck21}
{Ade}, P.~A.~R., {Ahmed}, Z., {Amiri}, M., {et~al.} 2021, \prl, 127, 151301

\bibitem[{{Aiola} {et~al.}(2020){Aiola}, {Calabrese}, {Maurin}, {Naess}, {Schmitt}, {Abitbol}, {Addison}, {Ade}, {Alonso}, {Amiri}, {Amodeo}, {Angile}, {Austermann}, {Baildon}, {Battaglia}, {Beall}, {Bean}, {Becker}, {Bond}, {Bruno}, {Calafut}, {Campusano}, {Carrero}, {Chesmore}, {Cho}, {Choi}, {Clark}, {Cothard}, {Crichton}, {Crowley}, {Darwish}, {Datta}, {Denison}, {Devlin}, {Duell}, {Duff}, {Duivenvoorden}, {Dunkley}, {D{\"u}nner}, {Essinger-Hileman}, {Fankhanel}, {Ferraro}, {Fox}, {Fuzia}, {Gallardo}, {Gluscevic}, {Golec}, {Grace}, {Gralla}, {Guan}, {Hall}, {Halpern}, {Han}, {Hargrave}, {Hasselfield}, {Helton}, {Henderson}, {Hensley}, {Hill}, {Hilton}, {Hilton}, {Hincks}, {Hlo{\v{z}}ek}, {Ho}, {Hubmayr}, {Huffenberger}, {Hughes}, {Infante}, {Irwin}, {Jackson}, {Klein}, {Knowles}, {Koopman}, {Kosowsky}, {Lakey}, {Li}, {Li}, {Li}, {Lokken}, {Louis}, {Lungu}, {MacInnis}, {Madhavacheril}, {Maldonado}, {Mallaby-Kay}, {Marsden}, {McMahon}, {Menanteau}, {Moodley}, {Morton}, {Namikawa}, {Nati}, {Newburgh},
  {Nibarger}, {Nicola}, {Niemack}, {Nolta}, {Orlowski-Sherer}, {Page}, {Pappas}, {Partridge}, {Phakathi}, {Pisano}, {Prince}, {Puddu}, {Qu}, {Rivera}, {Robertson}, {Rojas}, {Salatino}, {Schaan}, {Schillaci}, {Sehgal}, {Sherwin}, {Sierra}, {Sievers}, {Sifon}, {Sikhosana}, {Simon}, {Spergel}, {Staggs}, {Stevens}, {Storer}, {Sunder}, {Switzer}, {Thorne}, {Thornton}, {Trac}, {Treu}, {Tucker}, {Vale}, {Van Engelen}, {Van Lanen}, {Vavagiakis}, {Wagoner}, {Wang}, {Ward}, {Wollack}, {Xu}, {Zago}, \& {Zhu}}]{aiola20}
{Aiola}, S., {Calabrese}, E., {Maurin}, L., {et~al.} 2020, \jcap, 2020, 047

\bibitem[{{Arutjunjan} {et~al.}(2022){Arutjunjan}, {Schaefer}, \& {Kreutz}}]{schaefer22}
{Arutjunjan}, R., {Schaefer}, B.~M., \& {Kreutz}, C. 2022, arXiv e-prints, arXiv:2211.03421

\bibitem[{{Balkenhol} {et~al.}(2023){Balkenhol}, {Dutcher}, {Spurio Mancini}, {Doussot}, {Benabed}, {Galli}, {Ade}, {Anderson}, {Ansarinejad}, {Archipley}, {Bender}, {Benson}, {Bianchini}, {Bleem}, {Bouchet}, {Bryant}, {Camphuis}, {Carlstrom}, {Cecil}, {Chang}, {Chaubal}, {Chichura}, {Chou}, {Coerver}, {Crawford}, {Cukierman}, {Daley}, {de Haan}, {Dibert}, {Dobbs}, {Everett}, {Feng}, {Ferguson}, {Foster}, {Gambrel}, {Gardner}, {Goeckner-Wald}, {Gualtieri}, {Guidi}, {Guns}, {Halverson}, {Hivon}, {Holder}, {Holzapfel}, {Hood}, {Huang}, {Knox}, {Korman}, {Kuo}, {Lee}, {Lowitz}, {Lu}, {Millea}, {Montgomery}, {Nakato}, {Natoli}, {Noble}, {Novosad}, {Omori}, {Padin}, {Pan}, {Paschos}, {Prabhu}, {Quan}, {Rahimi}, {Rahlin}, {Reichardt}, {Rouble}, {Ruhl}, {Schiappucci}, {Smecher}, {Sobrin}, {Stark}, {Stephen}, {Suzuki}, {Tandoi}, {Thompson}, {Thorne}, {Tucker}, {Umilta}, {Vieira}, {Wang}, {Whitehorn}, {Wu}, {Yefremenko}, {Young}, {Zebrowski}, \& {SPT-3G Collaboration}}]{balkenhol23}
{Balkenhol}, L., {Dutcher}, D., {Spurio Mancini}, A., {et~al.} 2023, \prd, 108, 023510

\bibitem[{{Balkenhol} \& {Reichardt}(2022)}]{balkenhol22}
{Balkenhol}, L. \& {Reichardt}, C.~L. 2022, \mnras, 512, 4394

\bibitem[{Bradbury {et~al.}(2018)Bradbury, Frostig, Hawkins, Johnson, Leary, Maclaurin, Necula, Paszke, Vander{P}las, Wanderman-{M}ilne, \& Zhang}]{jax18}
Bradbury, J., Frostig, R., Hawkins, P., {et~al.} 2018, {JAX}: composable transformations of {P}ython+{N}um{P}y programs

\bibitem[{{Calabrese} {et~al.}(2013){Calabrese}, {Hlozek}, {Battaglia}, {Battistelli}, {Bond}, {Chluba}, {Crichton}, {Das}, {Devlin}, {Dunkley}, {D{\"u}nner}, {Farhang}, {Gralla}, {Hajian}, {Halpern}, {Hasselfield}, {Hincks}, {Irwin}, {Kosowsky}, {Louis}, {Marriage}, {Moodley}, {Newburgh}, {Niemack}, {Nolta}, {Page}, {Sehgal}, {Sherwin}, {Sievers}, {Sif{\'o}n}, {Spergel}, {Staggs}, {Switzer}, \& {Wollack}}]{calabrese13}
{Calabrese}, E., {Hlozek}, R.~A., {Battaglia}, N., {et~al.} 2013, \prd, 87, 103012

\bibitem[{{Campagne} {et~al.}(2023){Campagne}, {Lanusse}, {Zuntz}, {Boucaud}, {Casas}, {Karamanis}, {Kirkby}, {Lanzieri}, {Peel}, \& {Li}}]{campagne23}
{Campagne}, J.-E., {Lanusse}, F., {Zuntz}, J., {et~al.} 2023, The Open Journal of Astrophysics, 6, 15

\bibitem[{{Carlstrom} {et~al.}(2011){Carlstrom}, {Ade}, {Aird}, {Benson}, {Bleem}, {Busetti}, {Chang}, {Chauvin}, {Cho}, {Crawford}, {Crites}, {Dobbs}, {Halverson}, {Heimsath}, {Holzapfel}, {Hrubes}, {Joy}, {Keisler}, {Lanting}, {Lee}, {Leitch}, {Leong}, {Lu}, {Lueker}, {Luong-van}, {McMahon}, {Mehl}, {Meyer}, {Mohr}, {Montroy}, {Padin}, {Plagge}, {Pryke}, {Ruhl}, {Schaffer}, {Schwan}, {Shirokoff}, {Spieler}, {Staniszewski}, {Stark}, {Tucker}, {Vanderlinde}, {Vieira}, \& {Williamson}}]{carlstrom11}
{Carlstrom}, J.~E., {Ade}, P.~A.~R., {Aird}, K.~A., {et~al.} 2011, \pasp, 123, 568

\bibitem[{{Choi} {et~al.}(2020){Choi}, {Hasselfield}, {Ho}, {Koopman}, {Lungu}, {Abitbol}, {Addison}, {Ade}, {Aiola}, {Alonso}, {Amiri}, {Amodeo}, {Angile}, {Austermann}, {Baildon}, {Battaglia}, {Beall}, {Bean}, {Becker}, {Bond}, {Bruno}, {Calabrese}, {Calafut}, {Campusano}, {Carrero}, {Chesmore}, {Cho}, {Clark}, {Cothard}, {Crichton}, {Crowley}, {Darwish}, {Datta}, {Denison}, {Devlin}, {Duell}, {Duff}, {Duivenvoorden}, {Dunkley}, {D{\"u}nner}, {Essinger-Hileman}, {Fankhanel}, {Ferraro}, {Fox}, {Fuzia}, {Gallardo}, {Gluscevic}, {Golec}, {Grace}, {Gralla}, {Guan}, {Hall}, {Halpern}, {Han}, {Hargrave}, {Henderson}, {Hensley}, {Hill}, {Hilton}, {Hilton}, {Hincks}, {Hlo{\v{z}}ek}, {Hubmayr}, {Huffenberger}, {Hughes}, {Infante}, {Irwin}, {Jackson}, {Klein}, {Knowles}, {Kosowsky}, {Lakey}, {Li}, {Li}, {Li}, {Lokken}, {Louis}, {MacInnis}, {Madhavacheril}, {Maldonado}, {Mallaby-Kay}, {Marsden}, {Maurin}, {McMahon}, {Menanteau}, {Moodley}, {Morton}, {Naess}, {Namikawa}, {Nati}, {Newburgh}, {Nibarger}, {Nicola},
  {Niemack}, {Nolta}, {Orlowski-Sherer}, {Page}, {Pappas}, {Partridge}, {Phakathi}, {Prince}, {Puddu}, {Qu}, {Rivera}, {Robertson}, {Rojas}, {Salatino}, {Schaan}, {Schillaci}, {Schmitt}, {Sehgal}, {Sherwin}, {Sierra}, {Sievers}, {Sifon}, {Sikhosana}, {Simon}, {Spergel}, {Staggs}, {Stevens}, {Storer}, {Sunder}, {Switzer}, {Thorne}, {Thornton}, {Trac}, {Treu}, {Tucker}, {Vale}, {Van Engelen}, {Van Lanen}, {Vavagiakis}, {Wagoner}, {Wang}, {Ward}, {Wollack}, {Xu}, {Zago}, \& {Zhu}}]{choi20}
{Choi}, S.~K., {Hasselfield}, M., {Ho}, S.-P.~P., {et~al.} 2020, \jcap, 2020, 045

\bibitem[{{CMB-S4 Collaboration} {et~al.}(2019){CMB-S4 Collaboration}, {Abazajian}, {Addison}, {Adshead}, {Ahmed}, {Allen}, {Alonso}, {Alvarez}, {Anderson}, {Arnold}, {Baccigalupi}, \& et~al.}]{abazajian19}
{CMB-S4 Collaboration}, {Abazajian}, K., {Addison}, G., {et~al.} 2019, arXiv e-prints, arXiv:1907.04473

\bibitem[{{Dodelson} \& {Schneider}(2013)}]{dodelson13}
{Dodelson}, S. \& {Schneider}, M.~D. 2013, \prd, 88, 063537

\bibitem[{{Dunkley} {et~al.}(2013){Dunkley}, {Calabrese}, {Sievers}, {Addison}, {Battaglia}, {Battistelli}, {Bond}, {Das}, {Devlin}, {D{\"u}nner}, {Fowler}, {Gralla}, {Hajian}, {Halpern}, {Hasselfield}, {Hincks}, {Hlozek}, {Hughes}, {Irwin}, {Kosowsky}, {Louis}, {Marriage}, {Marsden}, {Menanteau}, {Moodley}, {Niemack}, {Nolta}, {Page}, {Partridge}, {Sehgal}, {Spergel}, {Staggs}, {Switzer}, {Trac}, \& {Wollack}}]{dunkley13}
{Dunkley}, J., {Calabrese}, E., {Sievers}, J., {et~al.} 2013, \jcap, 7, 25

\bibitem[{{Dutcher} {et~al.}(2021){Dutcher}, {Balkenhol}, {Ade}, {Ahmed}, {Anderes}, {Anderson}, {Archipley}, {Avva}, {Aylor}, {Barry}, {Basu Thakur}, {Benabed}, {Bender}, {Benson}, {Bianchini}, {Bleem}, {Bouchet}, {Bryant}, {Byrum}, {Carlstrom}, {Carter}, {Cecil}, {Chang}, {Chaubal}, {Chen}, {Cho}, {Chou}, {Cliche}, {Crawford}, {Cukierman}, {Daley}, {de Haan}, {Denison}, {Dibert}, {Ding}, {Dobbs}, {Everett}, {Feng}, {Ferguson}, {Foster}, {Fu}, {Galli}, {Gambrel}, {Gardner}, {Goeckner-Wald}, {Gualtieri}, {Guns}, {Gupta}, {Guyser}, {Halverson}, {Harke-Hosemann}, {Harrington}, {Henning}, {Hilton}, {Hivon}, {Holder}, {Holzapfel}, {Hood}, {Howe}, {Huang}, {Irwin}, {Jeong}, {Jonas}, {Jones}, {Khaire}, {Knox}, {Kofman}, {Korman}, {Kubik}, {Kuhlmann}, {Kuo}, {Lee}, {Leitch}, {Lowitz}, {Lu}, {Meyer}, {Michalik}, {Millea}, {Montgomery}, {Nadolski}, {Natoli}, {Nguyen}, {Noble}, {Novosad}, {Omori}, {Padin}, {Pan}, {Paschos}, {Pearson}, {Posada}, {Prabhu}, {Quan}, {Raghunathan}, {Rahlin}, {Reichardt}, {Riebel}, {Riedel},
  {Rouble}, {Ruhl}, {Sayre}, {Schiappucci}, {Shirokoff}, {Smecher}, {Sobrin}, {Stark}, {Stephen}, {Story}, {Suzuki}, {Thompson}, {Thorne}, {Tucker}, {Umilta}, {Vale}, {Vanderlinde}, {Vieira}, {Wang}, {Whitehorn}, {Wu}, {Yefremenko}, {Yoon}, {Young}, \& {SPT-3G Collaboration}}]{dutcher21}
{Dutcher}, D., {Balkenhol}, L., {Ade}, P.~A.~R., {et~al.} 2021, \prd, 104, 022003

\bibitem[{{Ge} {et~al.}(2024){Ge}, {Millea}, {Camphuis}, {Daley}, {Huang}, {Omori}, {Quan}, {Anderes}, {Anderson}, {Ansarinejad}, {Archipley}, {Balkenhol}, {Benabed}, {Bender}, {Benson}, {Bianchini}, {Bleem}, {Bouchet}, {Bryant}, {Carlstrom}, {Chang}, {Chaubal}, {Chen}, {Chichura}, {Chokshi}, {Chou}, {Coerver}, {Crawford}, {de Haan}, {Dibert}, {Dobbs}, {Doohan}, {Doussot}, {Dutcher}, {Everett}, {Feng}, {Ferguson}, {Fichman}, {Foster}, {Galli}, {Gambrel}, {Gardner}, {Goeckner-Wald}, {Gualtieri}, {Guidi}, {Guns}, {Halverson}, {Hivon}, {Holder}, {Holzapfel}, {Hood}, {Howe}, {Hryciuk}, {K{\'e}ruzor{\'e}}, {Khalife}, {Knox}, {Korman}, {Kornoelje}, {Kuo}, {Lee}, {Levy}, {Lowitz}, {Lu}, {Maniyar}, {Martsen}, {Menanteau}, {Montgomery}, {Nakato}, {Natoli}, {Noble}, {Pan}, {Paschos}, {Phadke}, {Pollak}, {Prabhu}, {Rahimi}, {Rahlin}, {Reichardt}, {Riebel}, {Rouble}, {Ruhl}, {Schiappucci}, {Sobrin}, {Stark}, {Stephen}, {Tandoi}, {Thorne}, {Trendafilova}, {Umilta}, {Vieira}, {Vitrier}, {Wan}, {Whitehorn}, {Wu}, {Young},
  \& {Zebrowski}}]{ge24}
{Ge}, F., {Millea}, M., {Camphuis}, E., {et~al.} 2024, arXiv e-prints, arXiv:2411.06000

\bibitem[{{Hartlap} {et~al.}(2007){Hartlap}, {Simon}, \& {Schneider}}]{hartlap06}
{Hartlap}, J., {Simon}, P., \& {Schneider}, P. 2007, \aap, 464, 399

\bibitem[{{Heavens} {et~al.}(2014){Heavens}, {Seikel}, {Nord}, {Aich}, {Bouffanais}, {Bassett}, \& {Hobson}}]{heavens14}
{Heavens}, A.~F., {Seikel}, M., {Nord}, B.~D., {et~al.} 2014, \mnras, 445, 1687

\bibitem[{{Henning} {et~al.}(2018){Henning}, {Sayre}, {Reichardt}, {Ade}, {Anderson}, {Austermann}, {Beall}, {Bender}, {Benson}, {Bleem}, {Carlstrom}, {Chang}, {Chiang}, {Cho}, {Citron}, {Corbett Moran}, {Crawford}, {Crites}, {de Haan}, {Dobbs}, {Everett}, {Gallicchio}, {George}, {Gilbert}, {Halverson}, {Harrington}, {Hilton}, {Holder}, {Holzapfel}, {Hoover}, {Hou}, {Hrubes}, {Huang}, {Hubmayr}, {Irwin}, {Keisler}, {Knox}, {Lee}, {Leitch}, {Li}, {Lowitz}, {Manzotti}, {McMahon}, {Meyer}, {Mocanu}, {Montgomery}, {Nadolski}, {Natoli}, {Nibarger}, {Novosad}, {Padin}, {Pryke}, {Ruhl}, {Saliwanchik}, {Schaffer}, {Sievers}, {Smecher}, {Stark}, {Story}, {Tucker}, {Vanderlinde}, {Veach}, {Vieira}, {Wang}, {Whitehorn}, {Wu}, \& {Yefremenko}}]{henning18}
{Henning}, J.~W., {Sayre}, J.~T., {Reichardt}, C.~L., {et~al.} 2018, \apj, 852, 97

\bibitem[{{Hoffman} \& {Gelman}(2011)}]{hoffman11}
{Hoffman}, M.~D. \& {Gelman}, A. 2011, arXiv e-prints, arXiv:1111.4246

\bibitem[{Hunter(2007)}]{hunter07}
Hunter, J.~D. 2007, Computing In Science \& Engineering, 9, 90

\bibitem[{{Jeong} {et~al.}(2014){Jeong}, {Chluba}, {Dai}, {Kamionkowski}, \& {Wang}}]{jeong14}
{Jeong}, D., {Chluba}, J., {Dai}, L., {Kamionkowski}, M., \& {Wang}, X. 2014, \prd, 89, 023003

\bibitem[{Jones {et~al.}(2001)Jones, Oliphant, Peterson, {et~al.}}]{jones01}
Jones, E., Oliphant, T., Peterson, P., {et~al.} 2001, {SciPy}: Open source scientific tools for {Python}, [Online; accessed 2014-10-22]

\bibitem[{{Keating} {et~al.}(2003){Keating}, {Ade}, {Bock}, {Hivon}, {Holzapfel}, {Lange}, {Nguyen}, \& {Yoon}}]{keating03b}
{Keating}, B.~G., {Ade}, P.~A.~R., {Bock}, J.~J., {et~al.} 2003, in Polarimetry in Astronomy. Edited by Silvano Fineschi. Proceedings of the SPIE, Volume 4843., 284--295

\bibitem[{{Kosowsky}(2003)}]{kosowsky03}
{Kosowsky}, A. 2003, in the proceedings of the workshop on ``The Cosmic Microwave Background and its Polarization," New Astronomy Reviews, ed. S.~Hanany \& K.~A. Olive (Elsevier)

\bibitem[{{Lewis} {et~al.}(2000){Lewis}, {Challinor}, \& {Lasenby}}]{lewis00}
{Lewis}, A., {Challinor}, A., \& {Lasenby}, A. 2000, \apj, 538, 473

\bibitem[{{Madhavacheril} {et~al.}(2024){Madhavacheril}, {Qu}, {Sherwin}, {MacCrann}, {Li}, {Abril-Cabezas}, {Ade}, {Aiola}, {Alford}, {Amiri}, {Amodeo}, {An}, {Atkins}, {Austermann}, {Battaglia}, {Battistelli}, {Beall}, {Bean}, {Beringue}, {Bhandarkar}, {Biermann}, {Bolliet}, {Bond}, {Cai}, {Calabrese}, {Calafut}, {Capalbo}, {Carrero}, {Challinor}, {Chesmore}, {Cho}, {Choi}, {Clark}, {C{\'o}rdova Rosado}, {Cothard}, {Coughlin}, {Coulton}, {Crowley}, {Dalal}, {Darwish}, {Devlin}, {Dicker}, {Doze}, {Duell}, {Duff}, {Duivenvoorden}, {Dunkley}, {D{\"u}nner}, {Fanfani}, {Fankhanel}, {Farren}, {Ferraro}, {Freundt}, {Fuzia}, {Gallardo}, {Garrido}, {Givans}, {Gluscevic}, {Golec}, {Guan}, {Hall}, {Halpern}, {Han}, {Harrison}, {Hasselfield}, {Healy}, {Henderson}, {Hensley}, {Herv{\'\i}as-Caimapo}, {Hill}, {Hilton}, {Hilton}, {Hincks}, {Hlo{\v{z}}ek}, {Ho}, {Huber}, {Hubmayr}, {Huffenberger}, {Hughes}, {Irwin}, {Isopi}, {Jense}, {Keller}, {Kim}, {Knowles}, {Koopman}, {Kosowsky}, {Kramer}, {Kusiak}, {La Posta}, {Lague},
  {Lakey}, {Lee}, {Li}, {Limon}, {Lokken}, {Louis}, {Lungu}, {MacInnis}, {Maldonado}, {Maldonado}, {Mallaby-Kay}, {Marques}, {McMahon}, {Mehta}, {Menanteau}, {Moodley}, {Morris}, {Mroczkowski}, {Naess}, {Namikawa}, {Nati}, {Newburgh}, {Nicola}, {Niemack}, {Nolta}, {Orlowski-Scherer}, {Page}, {Pandey}, {Partridge}, {Prince}, {Puddu}, {Radiconi}, {Robertson}, {Rojas}, {Sakuma}, {Salatino}, {Schaan}, {Schmitt}, {Sehgal}, {Shaikh}, {Sierra}, {Sievers}, {Sif{\'o}n}, {Simon}, {Sonka}, {Spergel}, {Staggs}, {Storer}, {Switzer}, {Tampier}, {Thornton}, {Trac}, {Treu}, {Tucker}, {Ullom}, {Vale}, {Van Engelen}, {Van Lanen}, {van Marrewijk}, {Vargas}, {Vavagiakis}, {Wagoner}, {Wang}, {Wenzl}, {Wollack}, {Xu}, {Zago}, \& {Zheng}}]{madhavacheril23}
{Madhavacheril}, M.~S., {Qu}, F.~J., {Sherwin}, B.~D., {et~al.} 2024, \apj, 962, 113

\bibitem[{{Manzotti} {et~al.}(2014){Manzotti}, {Hu}, \& {Benoit-L{\'e}vy}}]{manzotti14}
{Manzotti}, A., {Hu}, W., \& {Benoit-L{\'e}vy}, A. 2014, \prd, 90, 023003

\bibitem[{{Mocanu} {et~al.}(2019){Mocanu}, {Crawford}, {Aylor}, {Benson}, {Bleem}, {Carlstrom}, {Chang}, {Cho}, {Chown}, {Crites}, {de Haan}, {Dobbs}, {Everett}, {George}, {Halverson}, {Harrington}, {Henning}, {Holder}, {Holzapfel}, {Hou}, {Hrubes}, {Knox}, {Lee}, {Luong-Van}, {Marrone}, {McMahon}, {Meyer}, {Millea}, {Mohr}, {Natoli}, {Omori}, {Padin}, {Pryke}, {Reichardt}, {Ruhl}, {Sayre}, {Schaffer}, {Shirokoff}, {Staniszewski}, {Stark}, {Story}, {Vanderlinde}, {Vieira}, {Williamson}, \& {Wu}}]{mocanu19}
{Mocanu}, L.~M., {Crawford}, T.~M., {Aylor}, K., {et~al.} 2019, \jcap, 2019, 038

\bibitem[{Nash(1984)}]{nash84}
Nash, S.~G. 1984, SIAM Journal on Numerical Analysis, 21, 770

\bibitem[{Nocedal \& Wright(2006)}]{nocedal06}
Nocedal, J. \& Wright, S.~J. 2006, Numerical Optimization (New York, NY: Springer New York)

\bibitem[{{Pan} {et~al.}(2023){Pan}, {Bianchini}, {Wu}, {Ade}, {Ahmed}, {Anderes}, {Anderson}, {Ansarinejad}, {Archipley}, {Aylor}, {Balkenhol}, {Barry}, {Basu Thakur}, {Benabed}, {Bender}, {Benson}, {Bleem}, {Bouchet}, {Bryant}, {Byrum}, {Camphuis}, {Carlstrom}, {Carter}, {Cecil}, {Chang}, {Chaubal}, {Chen}, {Chichura}, {Cho}, {Chou}, {Cliche}, {Coerver}, {Crawford}, {Cukierman}, {Daley}, {de Haan}, {Denison}, {Dibert}, {Ding}, {Dobbs}, {Doussot}, {Dutcher}, {Everett}, {Feng}, {Ferguson}, {Fichman}, {Foster}, {Fu}, {Galli}, {Gambrel}, {Gardner}, {Ge}, {Goeckner-Wald}, {Gualtieri}, {Guidi}, {Guns}, {Gupta}, {Halverson}, {Harke-Hosemann}, {Harrington}, {Henning}, {Hilton}, {Hivon}, {Holder}, {Holzapfel}, {Hood}, {Howe}, {Huang}, {Irwin}, {Jeong}, {Jonas}, {Jones}, {K{\'e}ruzor{\'e}}, {Khaire}, {Knox}, {Kofman}, {Korman}, {Kubik}, {Kuhlmann}, {Kuo}, {Lee}, {Leitch}, {Levy}, {Lowitz}, {Lu}, {Maniyar}, {Menanteau}, {Meyer}, {Michalik}, {Millea}, {Montgomery}, {Nadolski}, {Nakato}, {Natoli}, {Nguyen}, {Noble},
  {Novosad}, {Omori}, {Padin}, {Paschos}, {Pearson}, {Posada}, {Prabhu}, {Quan}, {Raghunathan}, {Rahimi}, {Rahlin}, {Reichardt}, {Riebel}, {Riedel}, {Ruhl}, {Sayre}, {Schiappucci}, {Shirokoff}, {Smecher}, {Sobrin}, {Stark}, {Stephen}, {Story}, {Suzuki}, {Takakura}, {Tandoi}, {Thompson}, {Thorne}, {Trendafilova}, {Tucker}, {Umilta}, {Vale}, {Vanderlinde}, {Vieira}, {Wang}, {Whitehorn}, {Yefremenko}, {Yoon}, {Young}, \& {Zebrowski}}]{pan23}
{Pan}, Z., {Bianchini}, F., {Wu}, W.~L.~K., {et~al.} 2023, \prd, 108, 122005

\bibitem[{{Percival} {et~al.}(2022){Percival}, {Friedrich}, {Sellentin}, \& {Heavens}}]{percival22}
{Percival}, W.~J., {Friedrich}, O., {Sellentin}, E., \& {Heavens}, A. 2022, \mnras, 510, 3207

\bibitem[{{Piras} \& {Spurio Mancini}(2023)}]{piras23}
{Piras}, D. \& {Spurio Mancini}, A. 2023, The Open Journal of Astrophysics, 6, 20

\bibitem[{{Planck Collaboration} {et~al.}(2020{\natexlab{a}}){Planck Collaboration}, {Aghanim}, {Akrami}, {Arroja}, {Ashdown}, {Aumont}, {Baccigalupi}, {Ballardini}, {Banday}, {Barreiro}, {Bartolo}, {Basak}, {Battye}, {Benabed}, {Bernard}, {Bersanelli}, {Bielewicz}, {Bock}, {Bond}, {Borrill}, {Bouchet}, {Boulanger}, {Bucher}, {Burigana}, {Butler}, {Calabrese}, {Cardoso}, {Carron}, {Casaponsa}, {Challinor}, {Chiang}, {Colombo}, {Combet}, {Contreras}, {Crill}, {Cuttaia}, {de Bernardis}, {de Zotti}, {Delabrouille}, {Delouis}, {D{\'e}sert}, {Di Valentino}, {Dickinson}, {Diego}, {Donzelli}, {Dor{\'e}}, {Douspis}, {Ducout}, {Dupac}, {Efstathiou}, {Elsner}, {En{\ss}lin}, {Eriksen}, {Falgarone}, {Fantaye}, {Fergusson}, {Fernandez-Cobos}, {Finelli}, {Forastieri}, {Frailis}, {Franceschi}, {Frolov}, {Galeotta}, {Galli}, {Ganga}, {G{\'e}nova-Santos}, {Gerbino}, {Ghosh}, {Gonz{\'a}lez-Nuevo}, {G{\'o}rski}, {Gratton}, {Gruppuso}, {Gudmundsson}, {Hamann}, {Hand ley}, {Hansen}, {Helou}, {Herranz}, {Hildebrandt}, {Hivon},
  {Huang}, {Jaffe}, {Jones}, {Karakci}, {Keih{\"a}nen}, {Keskitalo}, {Kiiveri}, {Kim}, {Kisner}, {Knox}, {Krachmalnicoff}, {Kunz}, {Kurki-Suonio}, {Lagache}, {Lamarre}, {Langer}, {Lasenby}, {Lattanzi}, {Lawrence}, {Le Jeune}, {Leahy}, {Lesgourgues}, {Levrier}, {Lewis}, {Liguori}, {Lilje}, {Lilley}, {Lindholm}, {L{\'o}pez-Caniego}, {Lubin}, {Ma}, {Mac{\'\i}as-P{\'e}rez}, {Maggio}, {Maino}, {Mand olesi}, {Mangilli}, {Marcos-Caballero}, {Maris}, {Martin}, {Martinelli}, {Mart{\'\i}nez-Gonz{\'a}lez}, {Matarrese}, {Mauri}, {McEwen}, {Meerburg}, {Meinhold}, {Melchiorri}, {Mennella}, {Migliaccio}, {Millea}, {Mitra}, {Miville-Desch{\^e}nes}, {Molinari}, {Moneti}, {Montier}, {Morgante}, {Moss}, {Mottet}, {M{\"u}nchmeyer}, {Natoli}, {N{\o}rgaard-Nielsen}, {Oxborrow}, {Pagano}, {Paoletti}, {Partridge}, {Patanchon}, {Pearson}, {Peel}, {Peiris}, {Perrotta}, {Pettorino}, {Piacentini}, {Polastri}, {Polenta}, {Puget}, {Rachen}, {Reinecke}, {Remazeilles}, {Renault}, {Renzi}, {Rocha}, {Rosset}, {Roudier},
  {Rubi{\~n}o-Mart{\'\i}n}, {Ruiz-Granados}, {Salvati}, {Sandri}, {Savelainen}, {Scott}, {Shellard}, {Shiraishi}, {Sirignano}, {Sirri}, {Spencer}, {Sunyaev}, {Suur-Uski}, {Tauber}, {Tavagnacco}, {Tenti}, {Terenzi}, {Toffolatti}, {Tomasi}, {Trombetti}, {Valiviita}, {Van Tent}, {Vibert}, {Vielva}, {Villa}, {Vittorio}, {Wand elt}, {Wehus}, {White}, {White}, {Zacchei}, \& {Zonca}}]{planck18-1}
{Planck Collaboration}, {Aghanim}, N., {Akrami}, Y., {et~al.} 2020{\natexlab{a}}, \aap, 641, A1

\bibitem[{{Planck Collaboration} {et~al.}(2020{\natexlab{b}}){Planck Collaboration}, {Aghanim}, {Akrami}, {Ashdown}, {Aumont}, {Baccigalupi}, {Ballardini}, {Banday}, {Barreiro}, {Bartolo}, {Basak}, {Battye}, {Benabed}, {Bernard}, {Bersanelli}, {Bielewicz}, {Bock}, {Bond}, {Borrill}, {Bouchet}, {Boulanger}, {Bucher}, {Burigana}, {Butler}, {Calabrese}, {Cardoso}, {Carron}, {Challinor}, {Chiang}, {Chluba}, {Colombo}, {Combet}, {Contreras}, {Crill}, {Cuttaia}, {de Bernardis}, {de Zotti}, {Delabrouille}, {Delouis}, {Di Valentino}, {Diego}, {Dor{\'e}}, {Douspis}, {Ducout}, {Dupac}, {Dusini}, {Efstathiou}, {Elsner}, {En{\ss}lin}, {Eriksen}, {Fantaye}, {Farhang}, {Fergusson}, {Fernandez-Cobos}, {Finelli}, {Forastieri}, {Frailis}, {Fraisse}, {Franceschi}, {Frolov}, {Galeotta}, {Galli}, {Ganga}, {G{\'e}nova-Santos}, {Gerbino}, {Ghosh}, {Gonz{\'a}lez-Nuevo}, {G{\'o}rski}, {Gratton}, {Gruppuso}, {Gudmundsson}, {Hamann}, {Handley}, {Hansen}, {Herranz}, {Hildebrandt}, {Hivon}, {Huang}, {Jaffe}, {Jones}, {Karakci},
  {Keih{\"a}nen}, {Keskitalo}, {Kiiveri}, {Kim}, {Kisner}, {Knox}, {Krachmalnicoff}, {Kunz}, {Kurki-Suonio}, {Lagache}, {Lamarre}, {Lasenby}, {Lattanzi}, {Lawrence}, {Le Jeune}, {Lemos}, {Lesgourgues}, {Levrier}, {Lewis}, {Liguori}, {Lilje}, {Lilley}, {Lindholm}, {L{\'o}pez-Caniego}, {Lubin}, {Ma}, {Mac{\'\i}as-P{\'e}rez}, {Maggio}, {Maino}, {Mandolesi}, {Mangilli}, {Marcos-Caballero}, {Maris}, {Martin}, {Martinelli}, {Mart{\'\i}nez-Gonz{\'a}lez}, {Matarrese}, {Mauri}, {McEwen}, {Meinhold}, {Melchiorri}, {Mennella}, {Migliaccio}, {Millea}, {Mitra}, {Miville-Desch{\^e}nes}, {Molinari}, {Montier}, {Morgante}, {Moss}, {Natoli}, {N{\o}rgaard-Nielsen}, {Pagano}, {Paoletti}, {Partridge}, {Patanchon}, {Peiris}, {Perrotta}, {Pettorino}, {Piacentini}, {Polastri}, {Polenta}, {Puget}, {Rachen}, {Reinecke}, {Remazeilles}, {Renzi}, {Rocha}, {Rosset}, {Roudier}, {Rubi{\~n}o-Mart{\'\i}n}, {Ruiz-Granados}, {Salvati}, {Sandri}, {Savelainen}, {Scott}, {Shellard}, {Sirignano}, {Sirri}, {Spencer}, {Sunyaev}, {Suur-Uski},
  {Tauber}, {Tavagnacco}, {Tenti}, {Toffolatti}, {Tomasi}, {Trombetti}, {Valenziano}, {Valiviita}, {Van Tent}, {Vibert}, {Vielva}, {Villa}, {Vittorio}, {Wand elt}, {Wehus}, {White}, {White}, {Zacchei}, \& {Zonca}}]{planck18-6}
{Planck Collaboration}, {Aghanim}, N., {Akrami}, Y., {et~al.} 2020{\natexlab{b}}, \aap, 641, A6

\bibitem[{{Planck Collaboration} {et~al.}(2020{\natexlab{c}}){Planck Collaboration}, {Aghanim}, {Akrami}, {Ashdown}, {Aumont}, {Baccigalupi}, {Ballardini}, {Banday}, {Barreiro}, {Bartolo}, {Basak}, {Benabed}, {Bernard}, {Bersanelli}, {Bielewicz}, {Bock}, {Bond}, {Borrill}, {Bouchet}, {Boulanger}, {Bucher}, {Burigana}, {Butler}, {Calabrese}, {Cardoso}, {Carron}, {Casaponsa}, {Challinor}, {Chiang}, {Colombo}, {Combet}, {Crill}, {Cuttaia}, {de Bernardis}, {de Rosa}, {de Zotti}, {Delabrouille}, {Delouis}, {Di Valentino}, {Diego}, {Dor{\'e}}, {Douspis}, {Ducout}, {Dupac}, {Dusini}, {Efstathiou}, {Elsner}, {En{\ss}lin}, {Eriksen}, {Fantaye}, {Fernand ez-Cobos}, {Finelli}, {Frailis}, {Fraisse}, {Franceschi}, {Frolov}, {Galeotta}, {Galli}, {Ganga}, {G{\'e}nova-Santos}, {Gerbino}, {Ghosh}, {Giraud-H{\'e}raud}, {Gonz{\'a}lez-Nuevo}, {G{\'o}rski}, {Gratton}, {Gruppuso}, {Gudmundsson}, {Hamann}, {Handley}, {Hansen}, {Herranz}, {Hivon}, {Huang}, {Jaffe}, {Jones}, {Keih{\"a}nen}, {Keskitalo}, {Kiiveri}, {Kim}, {Kisner},
  {Krachmalnicoff}, {Kunz}, {Kurki-Suonio}, {Lagache}, {Lamarre}, {Lasenby}, {Lattanzi}, {Lawrence}, {Le Jeune}, {Levrier}, {Lewis}, {Liguori}, {Lilje}, {Lilley}, {Lindholm}, {L{\'o}pez-Caniego}, {Lubin}, {Ma}, {Mac{\'\i}as-P{\'e}rez}, {Maggio}, {Maino}, {Mandolesi}, {Mangilli}, {Marcos-Caballero}, {Maris}, {Martin}, {Mart{\'\i}nez-Gonz{\'a}lez}, {Matarrese}, {Mauri}, {McEwen}, {Meinhold}, {Melchiorri}, {Mennella}, {Migliaccio}, {Millea}, {Miville-Desch{\^e}nes}, {Molinari}, {Moneti}, {Montier}, {Morgante}, {Moss}, {Natoli}, {N{\o}rgaard-Nielsen}, {Pagano}, {Paoletti}, {Partridge}, {Patanchon}, {Peiris}, {Perrotta}, {Pettorino}, {Piacentini}, {Polenta}, {Puget}, {Rachen}, {Reinecke}, {Remazeilles}, {Renzi}, {Rocha}, {Rosset}, {Roudier}, {Rubi{\~n}o-Mart{\'\i}n}, {Ruiz-Granados}, {Salvati}, {Sandri}, {Savelainen}, {Scott}, {Shellard}, {Sirignano}, {Sirri}, {Spencer}, {Sunyaev}, {Suur-Uski}, {Tauber}, {Tavagnacco}, {Tenti}, {Toffolatti}, {Tomasi}, {Trombetti}, {Valiviita}, {Van Tent}, {Vielva}, {Villa},
  {Vittorio}, {Wandelt}, {Wehus}, {Zacchei}, \& {Zonca}}]{planck18-5}
{Planck Collaboration}, {Aghanim}, N., {Akrami}, Y., {et~al.} 2020{\natexlab{c}}, \aap, 641, A5

\bibitem[{{Planck Collaboration} {et~al.}(2016){Planck Collaboration}, {Aghanim}, {Arnaud}, {Ashdown}, {Aumont}, {Baccigalupi}, {Banday}, {Barreiro}, {Bartlett}, {Bartolo}, \& et~al.}]{planck15-11}
{Planck Collaboration}, {Aghanim}, N., {Arnaud}, M., {et~al.} 2016, \aap, 594, A11

\bibitem[{{Prince} {et~al.}(2024){Prince}, {Calabrese}, \& {Dunkley}}]{prince24}
{Prince}, H., {Calabrese}, E., \& {Dunkley}, J. 2024, arXiv e-prints, arXiv:2403.00085

\bibitem[{{Qu} {et~al.}(2024){Qu}, {Sherwin}, {Madhavacheril}, {Han}, {Crowley}, {Abril-Cabezas}, {Ade}, {Aiola}, {Alford}, {Amiri}, {Amodeo}, {An}, {Atkins}, {Austermann}, {Battaglia}, {Battistelli}, {Beall}, {Bean}, {Beringue}, {Bhandarkar}, {Biermann}, {Bolliet}, {Bond}, {Cai}, {Calabrese}, {Calafut}, {Capalbo}, {Carrero}, {Carron}, {Challinor}, {Chesmore}, {Cho}, {Choi}, {Clark}, {C{\'o}rdova Rosado}, {Cothard}, {Coughlin}, {Coulton}, {Dalal}, {Darwish}, {Devlin}, {Dicker}, {Doze}, {Duell}, {Duff}, {Duivenvoorden}, {Dunkley}, {D{\"u}nner}, {Fanfani}, {Fankhanel}, {Farren}, {Ferraro}, {Freundt}, {Fuzia}, {Gallardo}, {Garrido}, {Gluscevic}, {Golec}, {Guan}, {Halpern}, {Harrison}, {Hasselfield}, {Healy}, {Henderson}, {Hensley}, {Herv{\'\i}as-Caimapo}, {Hill}, {Hilton}, {Hilton}, {Hincks}, {Hlo{\v{z}}ek}, {Ho}, {Huber}, {Hubmayr}, {Huffenberger}, {Hughes}, {Irwin}, {Isopi}, {Jense}, {Keller}, {Kim}, {Knowles}, {Koopman}, {Kosowsky}, {Kramer}, {Kusiak}, {La Posta}, {Lague}, {Lakey}, {Lee}, {Li}, {Li}, {Limon},
  {Lokken}, {Louis}, {Lungu}, {MacCrann}, {MacInnis}, {Maldonado}, {Maldonado}, {Mallaby-Kay}, {Marques}, {McMahon}, {Mehta}, {Menanteau}, {Moodley}, {Morris}, {Mroczkowski}, {Naess}, {Namikawa}, {Nati}, {Newburgh}, {Nicola}, {Niemack}, {Nolta}, {Orlowski-Scherer}, {Page}, {Pandey}, {Partridge}, {Prince}, {Puddu}, {Radiconi}, {Robertson}, {Rojas}, {Sakuma}, {Salatino}, {Schaan}, {Schmitt}, {Sehgal}, {Shaikh}, {Sierra}, {Sievers}, {Sif{\'o}n}, {Simon}, {Sonka}, {Spergel}, {Staggs}, {Storer}, {Switzer}, {Tampier}, {Thornton}, {Trac}, {Treu}, {Tucker}, {Ullom}, {Vale}, {Van Engelen}, {Van Lanen}, {van Marrewijk}, {Vargas}, {Vavagiakis}, {Wagoner}, {Wang}, {Wenzl}, {Wollack}, {Xu}, {Zago}, \& {Zheng}}]{qu23}
{Qu}, F.~J., {Sherwin}, B.~D., {Madhavacheril}, M.~S., {et~al.} 2024, \apj, 962, 112

\bibitem[{{Sellentin} {et~al.}(2014){Sellentin}, {Quartin}, \& {Amendola}}]{sellentin14}
{Sellentin}, E., {Quartin}, M., \& {Amendola}, L. 2014, \mnras, 441, 1831

\bibitem[{{Simons Observatory Collaboration} {et~al.}(2019){Simons Observatory Collaboration}, {Ade}, {Aguirre}, {Ahmed}, {Aiola}, {Ali}, {Alonso}, {Alvarez}, {Arnold}, {Ashton}, {Austermann}, {Awan}, {Baccigalupi}, {Baildon}, {Barron}, {Battaglia}, {Battye}, {Baxter}, {Bazarko}, {Beall}, {Bean}, {Beck}, {Beckman}, {Beringue}, {Bianchini}, {Boada}, {Boettger}, {Bond}, {Borrill}, {Brown}, {Bruno}, {Bryan}, {Calabrese}, {Calafut}, {Calisse}, {Carron}, {Challinor}, {Chesmore}, {Chinone}, {Chluba}, {Cho}, {Choi}, {Coppi}, {Cothard}, {Coughlin}, {Crichton}, {Crowley}, {Crowley}, {Cukierman}, {D'Ewart}, {D{\"u}nner}, {de Haan}, {Devlin}, {Dicker}, {Didier}, {Dobbs}, {Dober}, {Duell}, {Duff}, {Duivenvoorden}, {Dunkley}, {Dusatko}, {Errard}, {Fabbian}, {Feeney}, {Ferraro}, {Flux{\`a}}, {Freese}, {Frisch}, {Frolov}, {Fuller}, {Fuzia}, {Galitzki}, {Gallardo}, {Tomas Galvez Ghersi}, {Gao}, {Gawiser}, {Gerbino}, {Gluscevic}, {Goeckner-Wald}, {Golec}, {Gordon}, {Gralla}, {Green}, {Grigorian}, {Groh}, {Groppi}, {Guan},
  {Gudmundsson}, {Han}, {Hargrave}, {Hasegawa}, {Hasselfield}, {Hattori}, {Haynes}, {Hazumi}, {He}, {Healy}, {Henderson}, {Hervias-Caimapo}, {Hill}, {Hill}, {Hilton}, {Hilton}, {Hincks}, {Hinshaw}, {Hlo{\v{z}}ek}, {Ho}, {Ho}, {Howe}, {Huang}, {Hubmayr}, {Huffenberger}, {Hughes}, {Ijjas}, {Ikape}, {Irwin}, {Jaffe}, {Jain}, {Jeong}, {Kaneko}, {Karpel}, {Katayama}, {Keating}, {Kernasovskiy}, {Keskitalo}, {Kisner}, {Kiuchi}, {Klein}, {Knowles}, {Koopman}, {Kosowsky}, {Krachmalnicoff}, {Kuenstner}, {Kuo}, {Kusaka}, {Lashner}, {Lee}, {Lee}, {Leon}, {Leung}, {Lewis}, {Li}, {Li}, {Limon}, {Linder}, {Lopez-Caraballo}, {Louis}, {Lowry}, {Lungu}, {Madhavacheril}, {Mak}, {Maldonado}, {Mani}, {Mates}, {Matsuda}, {Maurin}, {Mauskopf}, {May}, {McCallum}, {McKenney}, {McMahon}, {Meerburg}, {Meyers}, {Miller}, {Mirmelstein}, {Moodley}, {Munchmeyer}, {Munson}, {Naess}, {Nati}, {Navaroli}, {Newburgh}, {Nguyen}, {Niemack}, {Nishino}, {Orlowski-Scherer}, {Page}, {Partridge}, {Peloton}, {Perrotta}, {Piccirillo}, {Pisano},
  {Poletti}, {Puddu}, {Puglisi}, {Raum}, {Reichardt}, { }, {Rephaeli}, {Riechers}, {Rojas}, {Roy}, {Sadeh}, {Sakurai}, {Salatino}, {Sathyanarayana Rao}, {Schaan}, {Schmittfull}, {Sehgal}, {Seibert}, {Seljak}, {Sherwin}, {Shimon}, {Sierra}, {Sievers}, {Sikhosana}, {Silva-Feaver}, {Simon}, {Sinclair}, {Siritanasak}, {Smith}, {Smith}, {Spergel}, {Staggs}, {Stein}, {Stevens}, {Stompor}, {Suzuki}, {Tajima}, {Takakura}, {Teply}, {Thomas}, {Thorne}, {Thornton}, {Trac}, {Tsai}, {Tucker}, {Ullom}, {Vagnozzi}, {van Engelen}, {Van Lanen}, {Van Winkle}, {Vavagiakis}, {Verg{\`e}s}, {Vissers}, {Wagoner}, {Walker}, {Ward}, {Westbrook}, {Whitehorn}, {Williams}, {Williams}, {Wollack}, {Xu}, {Yu}, {Yu}, {Zago}, {Zhang}, {Zhu}, \& {The Simons Observatory collaboration}}]{ade19}
{Simons Observatory Collaboration}, {Ade}, P., {Aguirre}, J., {et~al.} 2019, \jcap, 2019, 056

\bibitem[{{Spurio Mancini} {et~al.}(2022){Spurio Mancini}, {Piras}, {Alsing}, {Joachimi}, \& {Hobson}}]{spuriomancini22}
{Spurio Mancini}, A., {Piras}, D., {Alsing}, J., {Joachimi}, B., \& {Hobson}, M.~P. 2022, \mnras, 511, 1771

\bibitem[{{Staniszewski} {et~al.}(2012){Staniszewski}, {Aikin}, {Amiri}, {Benton}, {Bischoff}, {Bock}, {Bonetti}, {Brevik}, {Burger}, {Dowell}, {Duband}, {Filippini}, {Golwala}, {Halpern}, {Hasselfield}, {Hilton}, {Hristov}, {Irwin}, {Kovac}, {Kuo}, {Lueker}, {Montroy}, {Nguyen}, {Ogburn}, {O'Brient}, {Orlando}, {Pryke}, {Reintsema}, {Ruhl}, {Schwarz}, {Sheehy}, {Stokes}, {Thompson}, {Teply}, {Tolan}, {Turner}, {Vieregg}, {Wilson}, {Wiebe}, \& {Wong}}]{staniszewski12}
{Staniszewski}, Z., {Aikin}, R.~W., {Amiri}, M., {et~al.} 2012, Journal of Low Temperature Physics, 167, 827

\bibitem[{{Torrado} \& {Lewis}(2021)}]{torrado21}
{Torrado}, J. \& {Lewis}, A. 2021, \jcap, 2021, 057

\bibitem[{van~der Walt {et~al.}(2011)van~der Walt, Colbert, \& Varoquaux}]{vanDerWalt11}
van~der Walt, S., Colbert, S., \& Varoquaux, G. 2011, Computing in Science Engineering, 13, 22

\bibitem[{{Virtanen} {et~al.}(2020){Virtanen}, {Gommers}, {Oliphant}, {Haberland}, {Reddy}, {Cournapeau}, {Burovski}, {Peterson}, {Weckesser}, {Bright}, {van der Walt}, {Brett}, {Wilson}, {Jarrod Millman}, {Mayorov}, {Nelson}, {Jones}, {Kern}, {Larson}, {Carey}, {Polat}, {Feng}, {Moore}, {Vand erPlas}, {Laxalde}, {Perktold}, {Cimrman}, {Henriksen}, {Quintero}, {Harris}, {Archibald}, {Ribeiro}, {Pedregosa}, {van Mulbregt}, \& {Contributors}}]{scipy}
{Virtanen}, P., {Gommers}, R., {Oliphant}, T.~E., {et~al.} 2020, Nature Methods, 17, 261

\end{thebibliography}

%%%%%%%%%%%%%%%%%%%%%%%%%%%%%%%%%%%%%%%%%%%%%%%%%%

\end{document}